\documentclass[aps,prap,amsmath,amssymb,reprint]{revtex4-1}

\usepackage{graphicx}
\usepackage{dcolumn}
\usepackage{bm}
\usepackage{lipsum}
\usepackage{appendix}

\begin{document}



\title{Thermometry of Silicon Nanoparticles}

\author{Matthew Mecklenburg}\email{matthew.mecklenburg@usc.edu }
\affiliation{Center for Electron Microscopy and Microanalysis, University of Southern California, Los Angeles,California 90089, USA}
\author{Brian Zutter}
\affiliation{Department of Physics \& Astronomy and California NanoSystems Institute, University of California, Los Angeles, California 90095 USA }
\author{B. C. Regan}\email{regan@physics.ucla.edu}
\affiliation{Department of Physics \& Astronomy and California NanoSystems Institute, University of California, Los Angeles, California 90095 USA }

\date{\today}

\begin{abstract}
Current thermometry techniques lack the spatial resolution required to see the temperature gradients in typical, highly-scaled modern transistors.  As a step toward addressing this problem, we have measured the temperature dependence of the volume plasmon energy in silicon nanoparticles from room temperature to 1250$^\circ$C, using a chip-style heating sample holder in a scanning transmission electron microscope (STEM) equipped with electron energy loss spectroscopy (EELS).  The plasmon energy changes as expected for an electron gas subject to the thermal expansion of silicon.  Reversing this reasoning, we find that measurements of the plasmon energy provide an independent measure of the nanoparticle temperature consistent with that of the heater chip's macroscopic heater/thermometer to within the 5\% accuracy of the chip thermometer's calibration.  Thus silicon has the potential to provide its own, high-spatial-resolution thermometric readout signal via measurements of its volume plasmon energy.  Furthermore, nanoparticles in general can serve as convenient nanothermometers for \emph{in situ} electron microscopy experiments.

\end{abstract}

\pacs{Valid PACS appear here}
\maketitle

Silicon, as the primary constituent of most semiconductor devices,  is perhaps the most important and most studied material in modern technology.  Silicon's thermal properties are particularly relevant to the design of devices such as microprocessors, since heat transport is frequently a performance-limiting factor in highly-scaled and high-power density electronics\cite{2016Heiderhoff,2014Cahill}. The current semiconductor processing node, designated with the scale label `10-nm', produces devices with features that are even smaller (in the vertical direction) and multiple, non-trivial interfaces.  

As such devices approach the atomic limit, classical, continuum thermal transport theory breaks down\cite{2016Heiderhoff,1938Casimir}. Improved designs for next-generation microprocessors, memory, and opto-electronics will come with a better understanding of thermal transport at these small length scales.  To gain this understanding,  thermometry techniques with $\lesssim 1\,\mu$m spatial resolution are required. However, no currently available technique can resolve the thermal gradients within the smallest modern transistors.

The temperature mapping techniques of most relevance to microelectronics are generally either optical or scanning-probe\cite{2016Heiderhoff,2015KimMicroscale}. Optical examples  include micro-Raman and thermoreflectance \cite{2014Maize,2008Beechem,1993Epperlein}, both of which are diffraction-limited to 500--1000~nm spatial resolution.  Mechanical scanning thermal microscopy (SThM) techniques do better by rastering a sharp tip across a sample\cite{1999Majumdar,2015Gomes}. They extract a thermometric signal by analyzing a tip-embedded thermometer\cite{1986Williams,1998Mills,2012KimUltra}, the heat transfer between the tip and sample\cite{1986Williams,2016MengesTemperature},  or the thermal expansion of the sample\cite{1998Majumdar}.

We are developing a temperature mapping technique, plasmon energy expansion thermometry (PEET) \cite{2015MecklenburgScience}, with the capability for $\lesssim 10$~nm spatial resolution inside a thermometric material. The technique is scanning, but, unlike most scanning techniques, it is non-contact in the sense that the  heat transfer between the probe and the sample is negligible. PEET infers a material's temperature from measurements of its volume plasmon energy.  The plasmon energy, $E_{p}=\hbar \sqrt{e^2 n/\epsilon_0 m}$ in the electron gas model (where $e$ and $m$ are the electronic charge and mass respectively), gives the valence electron density $n$.  The electron density in turn indicates the temperature via the material's coefficient of thermal expansion (CTE), which is determined separately.  In a scanning transmission electron microscope (STEM) equipped with electron energy loss spectroscopy (EELS), $E_p$ can be mapped with sufficiently high spatial resolution to observe the density changes at grain boundaries\cite{2015MecklenburgScience}.  Thus temperature mapping with resolution approaching the atomic limit can be achieved.

In this communication we share two main results.  First, we have measured the temperature dependence of silicon's bulk plasmon energy, which has not been reported previously.  This measurement is a necessary step toward the goal of applying PEET to determine the temperature gradients within an operating transistor, using the transistor's own silicon as the thermometric read-out material.  
 
Second, we show how nanoparticles can serve as fiducial thermometers for \emph{in situ} TEM experiments.  A compact PEET thermometer in or near the TEM field of view (FOV) can provide an improved temperature determination without the complications of external wiring or additional thermal loading.  Nanoparticles are small and can be easily dispersed.  With a variety of nanoparticles commercially available (\emph{e.g.} silicon, aluminum, indium, and tungsten), the specific type can be chosen to best meet the experiment's requirements (\emph{e.g.} operating temperature range and chemical compability).  Similar ideas for fiducial thermometers have been implemented previously in an optical context, for instance with nitrogen-vacancy centers in diamond \cite{2013Kucsko} or lanthanide ion-doped nanocrystals\cite{2010Vetrone}.  The PEET approach allows implementation in a TEM, and without requiring any additional hardware more exotic than a standard EELS spectrometer. In a sense each nanoparticle serves as an expansion thermometer in the style of Fahrenheit's mercury-in-glass design, but with a construction that is much simpler, cheaper, and smaller (vs., for example, the approach of Ref. \onlinecite{2002Gao}).

\begin{figure}
  \includegraphics[width=\columnwidth]{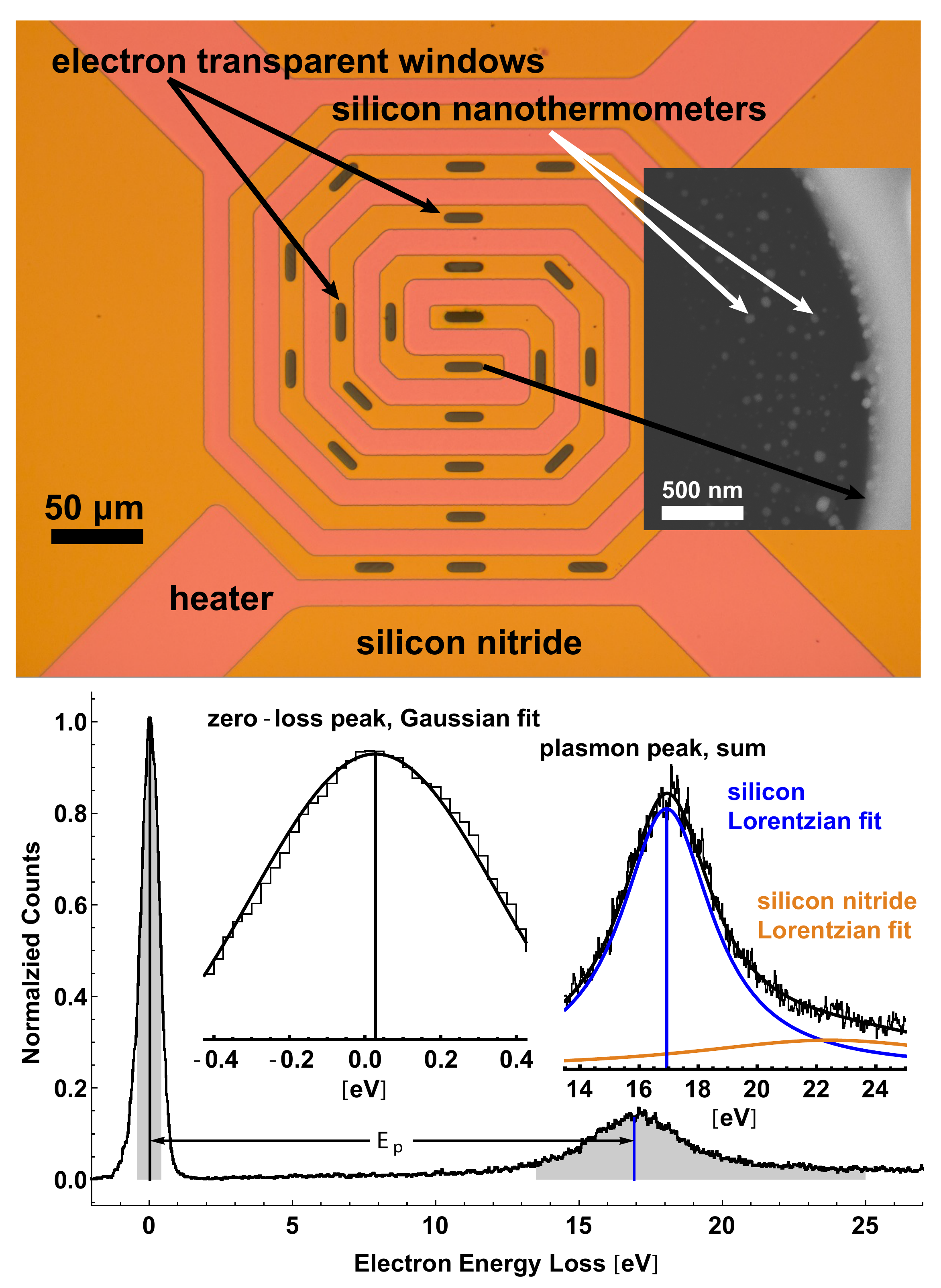}
  \caption{(\emph{top}) Chip-style TEM-sample heater. This optical micrograph shows the spiral heater and its four leads, which are used to make the resistance measurement that forms the basis of the chip's temperature determination.  At temperature the windows nearer the center of the spiral are hotter than those towards the edge, which emphasizes the desirability of having a small, local thermometer in the FOV. A scanning electron micrograph (inset) shows a typical dispersion of nanoparticles near the edge of one of the oblong, electron-transparent windows, and highlights the enormous size difference between these nanothermometers and the chip's dual-function heater/thermometer.  (\emph{bottom}) Low-loss EELS from a silicon nanoparticle.  The ZLP, silicon, and the silicon nitride plasmon peaks are fit to Gaussian, Lorentzian, and Lorentzian  functions respectively (insets) using data from the energy windows indicated by the grey vertical bands.}
	\label{fig:setup}
\end{figure}

To accomplish these two goals we measured the plasmon energy in silicon nanoparticles as a function of temperature using a chip-style TEM-sample heating holder (DENS Solutions Wildfire S3, Fig.~\ref{fig:setup} \emph{top}). Relative to  furnace-type heating holders, this type of holder equilibrates faster, drifts less, consumes less power, and  provides more accurate temperature read-out\cite{2016MeleMEMS}. As shown in Fig.~\ref{fig:setup} (\emph{top}), each chip had a $300\,\mu$m$\times 300\,\mu$m, SiN$_x$-encapsulated, spiral Joule heater/thermometer  atop a silicon nitride membrane with nearby $< 20$~nm-thick, $100\,\mu$m$^2$  electron-transparent  windows \cite{2016MeleMEMS,2012MeleMolybdenum}. The specifications for these chips list a guaranteed temperature range of room temperature to 1,300$^\circ$C, a maximum temperature of 1,500$^\circ$C, achievable temperature change rates of 200$^\circ$C/ms, and settling times of $<2$~s. At 1250$^\circ$C (1523~K) the heater drew 6.5~mA at 2.7~V, dissipating 18~mW. 

The window temperature was determined via a four-wire measurement of the heater resistance, which had been calibrated vs.\ temperature by the manufacturer to an accuracy of $5\%$.  By design the chip featured a temperature gradient, with the temperatures of different windows varying by more than $15$\%  relative to the difference from ambient at a given heater power.  The temperature calibration was only accurate for the windows nearest the center of the heater. 

Samples were prepared by dropcasting silicon nanoparticles from $1 \mu$l of an ethanol solution onto a chip (Figure 1).  According to the vendor (SkySpring Nanomaterials), the nanoparticles were manufactured by chemical vapor deposition (CVD), had  99\% purity, and a 100~nm average particle size. 

Generally speaking, 100~nm is roughly one mean-free path for plasmon production, so nanoparticles of this thickness are preferred for PEET.  Particle size-dependent effects are a potential source of systematic errors, but these only appear in much smaller particles. For instance, the bulk plasmon resonance has been observed to change in silicon nanoparticles with diameters $\lesssim 10$~nm\cite{1992Mitome}.  Similarly, size-dependent melting effects, which likely would have concomitant effects on the CTE, are only seen in particles with diameters $\lesssim 15$~nm\cite{2006Hirasawa}.

\begin{figure*}
  \includegraphics[width=\linewidth]{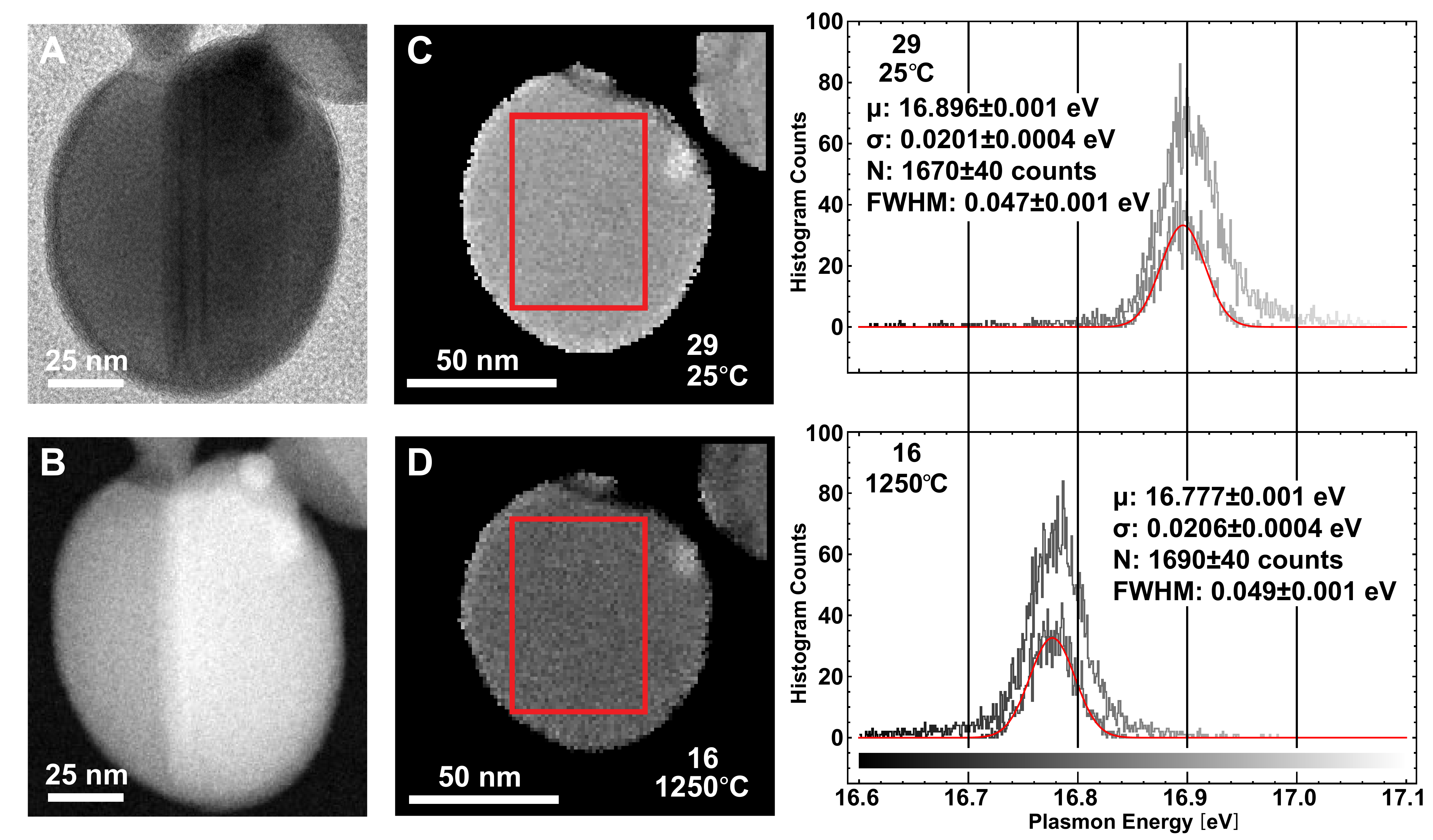}
  \caption{(A) TEM image of a silicon nanoparticle with at least two grains and an oxide coating. (B) Dark-field STEM image of the same nanoparticle. (C) and (D) Plasmon energy maps of the nanoparticle at 25$^\circ$C and 1250$^\circ$C respectively (the point number is listed above the temperature --- see Fig. 3). The combined scale bar/histograms to the right show  the distributions for the entire FOV, and the indicated red ROI.  The latter is fit to a Gaussian function.}
	\label{fig:Figure2}
\end{figure*}

EELS spectrum images of silicon nanoparticles at different temperatures were acquired in a JEOL JEM-2100F TEM equipped with a Gatan Quantum SE GIF.  The microscope was operated at 80~kV with a beam current of 100~pA, a 0.5~nm probe, and a convergence semi-angle of 12~mrads. (The 80~kV accelerating voltage enhances the plasmon production rate by roughly a factor of two relative to the rate at 200~kV.) The spectrometer collected 64 spectra per second with a semi-collection angle of 20 mrad, a 2.5~mm entrance aperture, a dispersion of 25~meV/channel, and $26\times$ vertical binning.

In each spectrum the silicon plasmon energy was determined by fitting the zero loss peak (ZLP), the silicon nitride plasmon peak, and the silicon  plasmon peak, as shown in Fig.~\ref{fig:setup}.  Fitting the ZLP with a Gaussian function in a fit window of full-width 0.85~$e$V centered around the spectrum maximum returned a full-width at half-maximum (FWHM) of $0.76\pm 0.01$~$e$V. In a region of interest (ROI) bare of any material but the electron-transparent membrane, the silicon nitride plasmon peak was fit with a Lorentzian function in a fit window extending from 19.5 to 26.5~$e$V relative to the ZLP center.  The peak center and width from this fit were then fixed, and a two-Lorentzian fit in the window 13.5--25.0~$e$V was performed over the entire FOV.  This fit had four free parameters: the amplitude of the silicon nitride peak, and the amplitude, center, and width of the silicon peak. The difference between the silicon peak center and the ZLP center is taken to be the silicon plasmon energy\cite{2015MecklenburgScience}. 

Typical data extracted from a 75~nm-diameter silicon nanoparticle are shown in Figure \ref{fig:Figure2}.  The TEM image with its diffraction contrast reveals the most detailed structural information, showing the nanoparticle's  8--10~nm-thick oxide coating and two distinct crystal grains.  The high-angle annular dark field STEM image shows the grains only, while the plasmon energy maps show none of these features and are basically uniform. Including the fit of the silicon nitride peak in the data analysis is necessary to achieve this uniformity; without it,  the plasmon energies within 10~nm of the nanoparticle edge  appear to be systematically higher than those in the interior (the low-amplitude silicon plasmon gets pulled higher by the slope in the silicon nitride background).  Histograms of the silicon plasmon energies are well-fit by Gaussian distributions. 

Converting these plasmon energy differences into temperature differences requires integrating silicon's linear CTE $\alpha(T) \equiv (1/l)(d l/d T)$, where $l$ is a length in the material\cite{1975Slack}.  The plasmon energies $E_p$ at an unknown temperature $T$ and the known reference temperature $T_0$ are related to the CTE by the ratio $R\equiv [{E_p(T)-E_p(T_0)}]/{E_p(T_0)}$, where 
\begin{equation}\label{first}
R\simeq -\frac{3}{2} \left(\frac{l-l_0}{l_0}\right)\simeq -\frac{3}{2}\int_{T_0}^T \alpha(T') dT'.
\end{equation}
Okada and Tokumaru\cite{1984Okada} provide an empirical formula for the CTE, valid between 120 and 1500~K, which integrated gives (for $T'$ in kelvin)
\begin{multline}\label{second}
\int \alpha(T') dT' =  
(1313.41 e^{-0.00588 T'} \\+ 3.725 T' + 0.0002774 T'^2)\times 10^{-6}.
\end{multline}
At $T= 300$, 600, 900, 1200, and 1500~K, this expression gives the CTE $\alpha = 2.57$, 3.83, 4.19, 4.38, and 4.56 (all $\times 10^{-6}$) respectively, which is to say that silicon's CTE is consistently increasing with temperature, though more slowly after a shoulder in the neighborhood of 700~K.  (Regarding PEET's sensitivity in silicon, it is unfortunate that, compared to that of other materials, silicon's high-temperature CTE is small, smaller even than that of diamond\cite{1975Slack}.)   In the range 298 to 1500~K  the integrated CTE $f(T)\equiv \int_{T_0}^T \alpha(T') dT'\simeq \alpha_1 \Delta T + \alpha_2 \Delta T^2$ ranges from 0 to $4.85\times 10^{-3}$ and is approximated with the coefficients $\alpha_1= 3.25\times 10^{-6}$~K$^{-1}$ and $\alpha_2=6.84\times 10^{-10}$~K$^{-2}$, where $\Delta T\equiv T-T_0$ and $T_0=298$~K. (For comparison, in aluminum the corresponding numbers\cite{2015MecklenburgScience} are $\alpha_1= 23.5\times 10^{-6}$~K$^{-1}$ and $\alpha_2= 89\times 10^{-10}$~K$^{-2}$ in the range 25 to 650$^\circ$C.)  However, while the quadratic approximation to Eq.~\ref{second} is good to better than $5\times10^{-5}$ through the whole range, the relative errors are as large as 27\% near room temperature where $f(T)$ is small.  Since for many applications the lower end of the range will be the most interesting region,  we invert $f(T)$ numerically to find temperatures.

Roughly speaking, silicon's plasmon shifts $-0.1$~m$e$V/K. Even a 1200~K temperature change produces a peak shift that is barely discernible by eye (see Supplementary Information). For the data in Fig.~\ref{fig:Figure2} the measured standard deviation of the single-pixel plasmon energies  is 20~m$e$V, which corresponds to a 200~K shift. With such uncertainties, meaningful temperatures cannot be calculated at the single-pixel level; the integrated CTE $f(T)$ is valid over only a limited temperature range. Furthermore,  $f(T)$ is non-linear.  Thus the operations of computing the temperature from the plasmon energies and averaging over some ROI do not commute ---  the averaging must be done first.  To suppress systematics arising from a weak silicon plasmon signal, we  compute the mean plasmon energy $\overline{E_p}(T)$ for an ROI in the interior of the nanoparticle at the unknown temperature $T$.  Finding the corresponding mean energy $\overline{E_p}(T_0)$ in a similar ROI in a map acquired at the reference temperature $T_0$, we calculate  $-2\overline{R}/3=f(T)$ and then invert to find the temperature.

\begin{figure}
  \includegraphics[width=0.9\columnwidth]{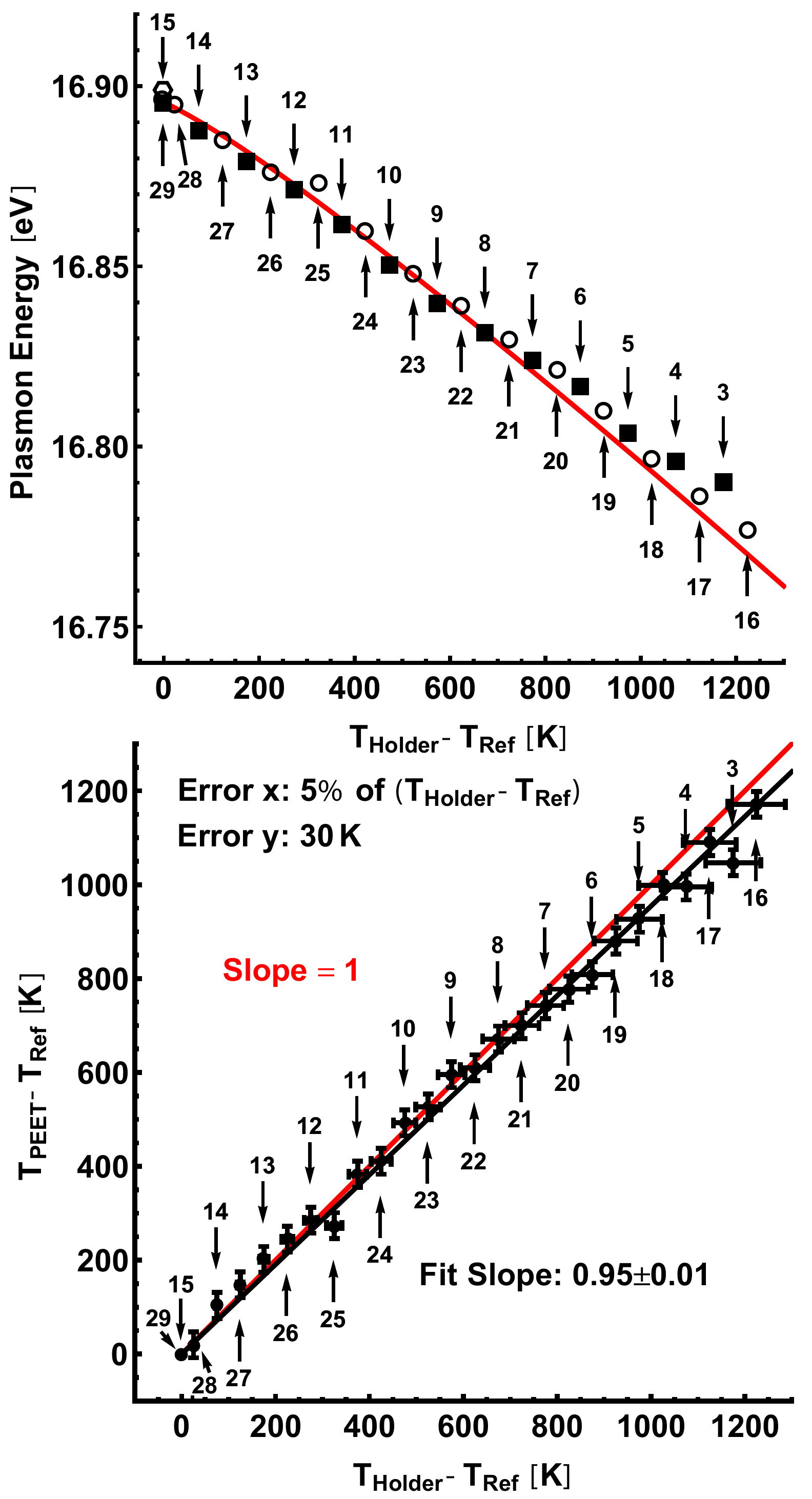}
  \caption{\emph{(top)}  The plasmon energy averaged over the ROI indicated in Fig.~2 is plotted as a function of the holder temperature.  Four ambient temperature measurements are shown (points labeled 1, 2, 15, and 29), along with two separate, high-to-low temperature ramps (black squares, 3--14, and open circles, 16--28, respectively). The measured plasmon energy changes follow the curve calculated using silicon's CTE. \emph{(bottom)} The corresponding PEET temperatures agree with the holder's temperature determination to within its stated 5\% accuracy.}
	\label{fig:Figure3}
\end{figure}

The nanoparticle plasmon energy maps shown in Fig.~\ref{fig:Figure2} represent two data points in a temperature scan designed to demonstrate the utility of such nanoparticles as nanothermometers. (For a more comprehensive view of the entire dataset see the Supplementary Information.)  This particular scan consisted of two room temperature data points, followed by two ramps down from high temperature to room temperature in 100$^\circ$C steps (according to temperature as determined by the holder), with the first ramp beginning at 1200$^\circ$C and the second at 1250$^\circ$C.  Interleaving two ramps with 100$^\circ$C steps, as opposed performing a single ramp with 50$^\circ$C steps, gives an important indication of the stability of the nanoparticles with respect to thermal cycling and repeated STEM imaging.  For maximum utility as nanothermometers, the nanoparticles should be robust to both perturbations.

The results of this scan are shown in Fig.~\ref{fig:Figure3}.  The plasmon energy versus temperature plot shows a total shift in the plasmon energy of 120~m$e$V --- a mere 3\%  of the peak's 3.7~$e$V  FWHM --- across the entire measured range between room temperature and 1523~K, highlighting the necessity of using curve-fitting to extract the thermometric signal. The plasmon energies determined in the two interleaved temperature ramps are themselves gratifyingly interleaved, showing no significant systematic shift between the first and second ramps.  To estimate the error in the PEET determination, we require that the $\chi^2$ per degree of freedom in the linear fit of Fig.~\ref{fig:Figure3} be unity, which gives a PEET error of 30~K. (Standard error propagation applied to the invertible, quadratic approximation to $f(T)$ gives errors that are too small by a factor of 8 for reasons that are not presently understood.) The four separate room-temperature plasmon energy measurements have a standard deviation of 2~m$e$V, an energy shift which is equivalent to 20~K. This value gives an additional measure of the error in PEET's temperature determination that is of the same order as the first.  Comparing the temperatures derived from resistance measurements of the chip's 300~$\mu$m heater/thermometer to those derived from PEET applied to the 75~nm silicon nanoparticle, we find that they agree at the 5\% level, the stated accuracy of the chip's temperature calibration. 

While applying PEET to nanoparticles we encountered various pitfalls, but the problems were usually easily recognized and even quantifiable. A change in the experimental parameters between the first and the last ambient-temperature measurements warns of a possible systematic. (Of course taking both measurements, and more within an experiment if possible, is a necessary part of a sound experimental protocol.) In cases with independent thermometers, like the one  described here, this warning might be unrelated to PEET and concern the other thermometer instead.  For instance, a change in the zero-power resistance of a heater/thermometer indicates that it has been damaged, either through use or through processing (\emph{e.g.} plasma cleaning), and that its temperature calibration can no longer be considered reliable.  In other cases the problem concerns PEET: the nanoparticle might change, either in its morphology, its plasmon energy, or both. We have seen evidence of alloying or doping within a heating experiment, and also signs of beam-induced damage.  Aberration-corrected microscopes are particularly hazardous in the latter regard, for a total beam current that is harmless in an uncorrected probe can, concentrated, radically transform a nanoparticle, making it useless for thermometry. Whatever the source of the change, the shift in a nanoparticle's plasmon energy under nominally identical conditions gives a quantitative measure of the magnitude of a potential systematic. 
\begin{acknowledgments}
This work was supported by FAME, one of six centers of STARnet, a Semiconductor Research Corporation program sponsored by MARCO and DARPA, by National Science Foundation (NSF) award DMR-1611036, and by NSF STC award DMR-1548924. The data presented were acquired at the Center for Electron Microscopy and Microanalysis at the University of Southern California.
\end{acknowledgments}


\bibliographystyle{naturemag9}
\bibliography{Al_Si_Plasmon_cr}

\begin{thebibliography}{10}
\expandafter\ifx\csname url\endcsname\relax
  \def\url#1{\texttt{#1}}\fi
\expandafter\ifx\csname urlprefix\endcsname\relax\def\urlprefix{URL }\fi
\providecommand{\bibinfo}[2]{#2}
\providecommand{\eprint}[2][]{\url{#2}}

\bibitem{2016Heiderhoff}
\bibinfo{author}{Heiderhoff, R.}, \bibinfo{author}{Makris, A.} \&
  \bibinfo{author}{Riedl, T.}
\newblock \bibinfo{title}{Thermal microscopy of electronic materials}.
\newblock \emph{\bibinfo{journal}{Materials Science in Semiconductor
  Processing}} \textbf{\bibinfo{volume}{43}}, \bibinfo{pages}{163--176}
  (\bibinfo{year}{2016}).

\bibitem{2014Cahill}
\bibinfo{author}{Cahill, D.~G.} \emph{et~al.}
\newblock \bibinfo{title}{Nanoscale thermal transport. {{II}}.
  2003\textendash{}2012}.
\newblock \emph{\bibinfo{journal}{Applied Physics Reviews}}
  \textbf{\bibinfo{volume}{1}}, \bibinfo{pages}{011305} (\bibinfo{year}{2014}).

\bibitem{1938Casimir}
\bibinfo{author}{Casimir, H. B.~G.}
\newblock \bibinfo{title}{Note on the conduction of heat in crystals}.
\newblock \emph{\bibinfo{journal}{Physica}} \textbf{\bibinfo{volume}{5}},
  \bibinfo{pages}{495--500} (\bibinfo{year}{1938}).

\bibitem{2015KimMicroscale}
\bibinfo{author}{Kim, M.~M.}, \bibinfo{author}{Giry, A.},
  \bibinfo{author}{Mastiani, M.}, \bibinfo{author}{Rodrigues, G.~O.},
  \bibinfo{author}{Reis, A.} \& \bibinfo{author}{Mandin, P.}
\newblock \bibinfo{title}{Microscale thermometry: {{A}} review}.
\newblock \emph{\bibinfo{journal}{Microelectronic Engineering}}
  \textbf{\bibinfo{volume}{148}}, \bibinfo{pages}{129--142}
  (\bibinfo{year}{2015}).

\bibitem{2014Maize}
\bibinfo{author}{Maize, K.}, \bibinfo{author}{Pavlidis, G.},
  \bibinfo{author}{Heller, E.}, \bibinfo{author}{Yates, L.},
  \bibinfo{author}{Kendig, D.}, \bibinfo{author}{Graham, S.} \&
  \bibinfo{author}{Shakouri, A.}
\newblock \bibinfo{title}{High {{Resolution Thermal Characterization}} and
  {{Simulation}} of {{Power AlGaN}}/{{GaN HEMTs Using Micro}}-{{Raman
  Thermography}} and 800 {{Picosecond Transient Thermoreflectance Imaging}}}.
\newblock In \emph{\bibinfo{booktitle}{2014 {{IEEE Compound Semiconductor
  Integrated Circuit Symposium}} ({{CSICs}})}}, \bibinfo{pages}{1--8}
  (\bibinfo{year}{2014}).

\bibitem{2008Beechem}
\bibinfo{author}{Beechem, T.}, \bibinfo{author}{Christensen, A.},
  \bibinfo{author}{Graham, S.} \& \bibinfo{author}{Green, D.}
\newblock \bibinfo{title}{Micro-{{Raman}} thermometry in the presence of
  complex stresses in {{GaN}} devices}.
\newblock \emph{\bibinfo{journal}{Journal of Applied Physics}}
  \textbf{\bibinfo{volume}{103}}, \bibinfo{pages}{124501}
  (\bibinfo{year}{2008}).

\bibitem{1993Epperlein}
\bibinfo{author}{Epperlein, P.-W.}
\newblock \bibinfo{title}{Micro-{{Temperature Measurements}} on {{Semiconductor
  Laser Mirrors}} by {{Reflectance Modulation}}: {{A Newly Developed
  Technique}} for {{Laser Characterization}}}.
\newblock \emph{\bibinfo{journal}{Japanese Journal of Applied Physics}}
  \textbf{\bibinfo{volume}{32}}, \bibinfo{pages}{5514--5522}
  (\bibinfo{year}{1993}).

\bibitem{1999Majumdar}
\bibinfo{author}{Majumdar, A.}
\newblock \bibinfo{title}{Scanning {{Thermal Microscopy}}}.
\newblock \emph{\bibinfo{journal}{Annual Review of Materials Science}}
  \textbf{\bibinfo{volume}{29}}, \bibinfo{pages}{505--585}
  (\bibinfo{year}{1999}).

\bibitem{2015Gomes}
\bibinfo{author}{Gom{\`e}s, S.}, \bibinfo{author}{Assy, A.} \&
  \bibinfo{author}{Chapuis, P.-O.}
\newblock \bibinfo{title}{Scanning thermal microscopy: {{A}} review}.
\newblock \emph{\bibinfo{journal}{physica status solidi (a)}}
  \textbf{\bibinfo{volume}{212}}, \bibinfo{pages}{477--494}
  (\bibinfo{year}{2015}).

\bibitem{1986Williams}
\bibinfo{author}{Williams, C.~C.} \& \bibinfo{author}{Wickramasinghe, H.~K.}
\newblock \bibinfo{title}{Scanning thermal profiler}.
\newblock \emph{\bibinfo{journal}{Applied Physics Letters}}
  \textbf{\bibinfo{volume}{49}}, \bibinfo{pages}{1587--1589}
  (\bibinfo{year}{1986}).

\bibitem{1998Mills}
\bibinfo{author}{Mills, G.}, \bibinfo{author}{Zhou, H.},
  \bibinfo{author}{Midha, A.}, \bibinfo{author}{Donaldson, L.} \&
  \bibinfo{author}{Weaver, J. M.~R.}
\newblock \bibinfo{title}{Scanning thermal microscopy using batch fabricated
  thermocouple probes}.
\newblock \emph{\bibinfo{journal}{Applied Physics Letters}}
  \textbf{\bibinfo{volume}{72}}, \bibinfo{pages}{2900--2902}
  (\bibinfo{year}{1998}).

\bibitem{2012KimUltra}
\bibinfo{author}{Kim, K.}, \bibinfo{author}{Jeong, W.}, \bibinfo{author}{Lee,
  W.} \& \bibinfo{author}{Reddy, P.}
\newblock \bibinfo{title}{Ultra-{{High Vacuum Scanning Thermal Microscopy}} for
  {{Nanometer Resolution Quantitative Thermometry}}}.
\newblock \emph{\bibinfo{journal}{ACS Nano}} \textbf{\bibinfo{volume}{6}},
  \bibinfo{pages}{4248--4257} (\bibinfo{year}{2012}).

\bibitem{2016MengesTemperature}
\bibinfo{author}{Menges, F.}, \bibinfo{author}{Mensch, P.},
  \bibinfo{author}{Schmid, H.}, \bibinfo{author}{Riel, H.},
  \bibinfo{author}{Stemmer, A.} \& \bibinfo{author}{Gotsmann, B.}
\newblock \bibinfo{title}{Temperature mapping of operating nanoscale devices by
  scanning probe thermometry}.
\newblock \emph{\bibinfo{journal}{Nature Communications}}
  \textbf{\bibinfo{volume}{7}}, \bibinfo{pages}{10874} (\bibinfo{year}{2016}).

\bibitem{1998Majumdar}
\bibinfo{author}{Majumdar, A.} \& \bibinfo{author}{Varesi, J.}
\newblock \bibinfo{title}{Nanoscale {{Temperature Distributions Measured}} by
  {{Scanning Joule Expansion Microscopy}}}.
\newblock \emph{\bibinfo{journal}{Journal of Heat Transfer}}
  \textbf{\bibinfo{volume}{120}}, \bibinfo{pages}{297--305}
  (\bibinfo{year}{1998}).

\bibitem{2015MecklenburgScience}
\bibinfo{author}{Mecklenburg, M.}, \bibinfo{author}{Hubbard, W.~A.},
  \bibinfo{author}{White, E.~R.}, \bibinfo{author}{Dhall, R.},
  \bibinfo{author}{Cronin, S.~B.}, \bibinfo{author}{Aloni, S.} \&
  \bibinfo{author}{Regan, B.~C.}
\newblock \bibinfo{title}{Nanoscale temperature mapping in operating
  microelectronic devices}.
\newblock \emph{\bibinfo{journal}{Science}} \textbf{\bibinfo{volume}{347}},
  \bibinfo{pages}{629--632} (\bibinfo{year}{2015}).

\bibitem{2013Kucsko}
\bibinfo{author}{Kucsko, G.}, \bibinfo{author}{Maurer, P.~C.},
  \bibinfo{author}{Yao, N.~Y.}, \bibinfo{author}{Kubo, M.},
  \bibinfo{author}{Noh, H.~J.}, \bibinfo{author}{Lo, P.~K.},
  \bibinfo{author}{Park, H.} \& \bibinfo{author}{Lukin, M.~D.}
\newblock \bibinfo{title}{Nanometre-scale thermometry in a living cell}.
\newblock \emph{\bibinfo{journal}{Nature}} \textbf{\bibinfo{volume}{500}},
  \bibinfo{pages}{54--58} (\bibinfo{year}{2013}).

\bibitem{2010Vetrone}
\bibinfo{author}{Vetrone, F.} \emph{et~al.}
\newblock \bibinfo{title}{Temperature {{Sensing Using Fluorescent
  Nanothermometers}}}.
\newblock \emph{\bibinfo{journal}{ACS Nano}} \textbf{\bibinfo{volume}{4}},
  \bibinfo{pages}{3254--3258} (\bibinfo{year}{2010}).

\bibitem{2002Gao}
\bibinfo{author}{Gao, Y.} \& \bibinfo{author}{Bando, Y.}
\newblock \bibinfo{title}{Nanotechnology: {{Carbon}} nanothermometer containing
  gallium}.
\newblock \emph{\bibinfo{journal}{Nature}} \textbf{\bibinfo{volume}{415}},
  \bibinfo{pages}{599--599} (\bibinfo{year}{2002}).

\bibitem{2016MeleMEMS}
\bibinfo{author}{Mele, L.}, \bibinfo{author}{Konings, S.},
  \bibinfo{author}{Dona, P.}, \bibinfo{author}{Evertz, F.},
  \bibinfo{author}{Mitterbauer, C.}, \bibinfo{author}{Faber, P.},
  \bibinfo{author}{Schampers, R.} \& \bibinfo{author}{Jinschek, J.~R.}
\newblock \bibinfo{title}{A {{MEMS}}-based heating holder for the direct
  imaging of simultaneous in-situ heating and biasing experiments in
  scanning/transmission electron microscopes}.
\newblock \emph{\bibinfo{journal}{Microscopy Research and Technique}}
  \textbf{\bibinfo{volume}{79}}, \bibinfo{pages}{239--250}
  (\bibinfo{year}{2016}).

\bibitem{2012MeleMolybdenum}
\bibinfo{author}{Mele, L.}, \bibinfo{author}{Santagata, F.},
  \bibinfo{author}{Iervolino, E.}, \bibinfo{author}{Mihailovic, M.},
  \bibinfo{author}{Rossi, T.}, \bibinfo{author}{Tran, A.~T.},
  \bibinfo{author}{Schellevis, H.}, \bibinfo{author}{Creemer, J.~F.} \&
  \bibinfo{author}{Sarro, P.~M.}
\newblock \bibinfo{title}{A molybdenum {{MEMS}} microhotplate for
  high-temperature operation}.
\newblock \emph{\bibinfo{journal}{Sensors and Actuators A: Physical}}
  \textbf{\bibinfo{volume}{188}}, \bibinfo{pages}{173--180}
  (\bibinfo{year}{2012}).

\bibitem{1992Mitome}
\bibinfo{author}{Mitome, M.}, \bibinfo{author}{Yamazaki, Y.},
  \bibinfo{author}{Takagi, H.} \& \bibinfo{author}{Nakagiri, T.}
\newblock \bibinfo{title}{Size dependence of plasmon energy in {{Si}}
  clusters}.
\newblock \emph{\bibinfo{journal}{Journal of Applied Physics}}
  \textbf{\bibinfo{volume}{72}}, \bibinfo{pages}{812--814}
  (\bibinfo{year}{1992}).

\bibitem{2006Hirasawa}
\bibinfo{author}{Hirasawa, M.}, \bibinfo{author}{Orii, T.} \&
  \bibinfo{author}{Seto, T.}
\newblock \bibinfo{title}{Size-dependent crystallization of {{Si}}
  nanoparticles}.
\newblock \emph{\bibinfo{journal}{Applied Physics Letters}}
  \textbf{\bibinfo{volume}{88}}, \bibinfo{pages}{093119}
  (\bibinfo{year}{2006}).

\bibitem{1975Slack}
\bibinfo{author}{Slack, G.~A.} \& \bibinfo{author}{Bartram, S.~F.}
\newblock \bibinfo{title}{Thermal expansion of some diamondlike crystals}.
\newblock \emph{\bibinfo{journal}{Journal of Applied Physics}}
  \textbf{\bibinfo{volume}{46}}, \bibinfo{pages}{89--98}
  (\bibinfo{year}{1975}).

\bibitem{1984Okada}
\bibinfo{author}{Okada, Y.} \& \bibinfo{author}{Tokumaru, Y.}
\newblock \bibinfo{title}{Precise determination of lattice parameter and
  thermal expansion coefficient of silicon between 300 and 1500 {{K}}}.
\newblock \emph{\bibinfo{journal}{Journal of Applied Physics}}
  \textbf{\bibinfo{volume}{56}}, \bibinfo{pages}{314--320}
  (\bibinfo{year}{1984}).

\end{thebibliography}

\end{document}